\documentclass[aps,prx,twocolumn,amsmath,amssymb]{revtex4-1}
\usepackage{graphicx}
\usepackage{dcolumn}
\usepackage{bm}
\usepackage{amsmath}
\usepackage{amsfonts}
\usepackage{amssymb}
\usepackage{amsfonts}
\usepackage{amssymb}
\usepackage{multirow}

\usepackage{amsmath}
\usepackage{amssymb}
\usepackage{amsthm}
\usepackage{pinlabel}
\usepackage{natbib}

\usepackage{epstopdf}
\usepackage[abs]{overpic}
\usepackage{dcolumn}

\usepackage{color}

\usepackage[dvipsnames]{xcolor}

\newcommand{\no}{\nonumber}

\usepackage{braket}

\definecolor{specialgray}{HTML}{505050}
\definecolor{col10K}{HTML}{FFA000}
\definecolor{col300K}{HTML}{924FA4}
\definecolor{colMu}{HTML}{5278BD}
\definecolor{colMuI}{HTML}{924FA4}

\begin{document}

\title{Increased Performance of Matsubara space calculations: A case study within Eliashberg theory}
\author{Fabian Schrodi}\email{fabian.schrodi@physics.uu.se}
\author{Alex Aperis}\email{alex.aperis@physics.uu.se}
\author{Peter M. Oppeneer}\email{peter.oppeneer@physics.uu.se}
\affiliation{Department of Physics and Astronomy, Uppsala University, P.\ O.\ Box 516, SE-75120 Uppsala, Sweden}

%\vskip 0.4cm
\date{\today}

\begin{abstract}
\noindent 
We present a method to considerably improve the numerical performance for solving Eliashberg-type coupled equations on the imaginary axis. Instead of the standard practice of introducing a hard numerical cutoff for treating the infinite summations involved, our scheme allows for the efficient calculation of such sums extended formally up to infinity. The method is first benchmarked with isotropic Migdal-Eliashberg theory calculations	and subsequently applied to the solution of the full-bandwidth, multiband and anisotropic equations focusing on the FeSe/SrTiO$_3$ interface as a case study. Compared to the standard procedure, we reach similarly well converged results with less than one fifth of the number of frequencies for the anisotropic case, while for the isotropic set of equations we spare approximately ninety percent of the complexity. Since our proposed approximations are very general, our numerical scheme opens the possibility of studying the superconducting properties of a wide range of materials at ultra-low temperatures.
\end{abstract}

\maketitle

\section{Introduction}

The most successful theory for explaining superconductivity in real materials is arguably Eliashberg theory \cite{eliashberg1960}, which generalizes the Bardeen-Cooper-Schrieffer (BCS) description of superconductors \cite{bardeen1957} by explicitly taking into account the retarded nature of the electron-boson interaction that mediates the Cooper pairing \cite{Allen1983,carbotte1990,bennemann2008}. The self-consistent solution of the Eliashberg equations, supplemented with \textit{ab initio} calculated input for the electrons and the involved bosons, has evolved into a powerful method for the materials specific modeling of superconductors on the quantitative level \cite{Choi2002,Margine2013,Aperis2015,bekaert2016,bekaert2018}. Yet, calculations of such fidelity remain to a large extent computationally expensive since they typically involve coupled integral equations over momentum and frequency.
Due to the concomitant computational complexity of the involved principal value integrals, the Eliashberg equations are rarely solved directly in the real frequency domain. Instead, they are commonly treated on the imaginary (Matsubara) frequency axis and, if needed, the real frequency dependence can be retrieved via a numerical analytic continuation \cite{Vidberg1977,marsiglio1988}.

Solving the self-consistent problem numerically is exceptionally hard in general, because there is a need of performing Matsubara sums over ideally infinitely many bosonic or fermionic frequencies. Although most often the physically interesting region is relatively small and centered around the zeroth frequency, the results can still heavily rely on the number of Matsubara frequencies involved in the calculations. The generally accepted procedure to address this challenge is to introduce a symmetric frequency interval via a hard numerical cutoff $\mathcal{M}$, and neglect all contributions outside of this interval \cite{Allen1983,carbotte1990}. The boundaries $\pm\mathcal{M}$ are symmetrically shifted until the results (e.g. the gap function) do not change significantly anymore; such convergence study makes the use of a large number of Matsubara frequencies a necessity. Apart from the risk of obtaining results that are not well converged due to a lack of numerical resources, the equations in Matsubara space provide another difficulty: The frequencies scale with temperature, which requires an even larger number of frequencies to account for the physically relevant energy window in low temperature systems.  This can potentially become a bottleneck for the efficiency of calculations where full momentum dependence needs to be retained, depending also on the specifics of a given material. Typically, such a situation may be encountered e.g. in fully anisotropic Eliashberg simulations with \textit{ab initio} input \cite{bekaert2016}, or when the electron energy dispersions throughout the full bandwidth are explicitly included in the calculation \cite{aperis2018,schrodi2018}. Therefore, as the need for more realistic Eliashberg calculations grows, improving the associated computational performance becomes a necessity. 

It is worth mentioning that the problem of efficiently treating infinite Matsubara sums is not peculiar to Eliashberg theory-type of calculations but occurs rather often in numerical solutions of Dyson-like self-consistent equations that involve convolutions between propagators. For example, calculations within the fluctuation-exchange (FLEX) approximation require the knowledge of susceptibilities that can be numerically calculated as convolutions between two Green's functions. For that particular case, it has been proposed that high frequency corrections can be obtained by improving the analytical properties of the propagator in the imaginary time domain via the so-called $\tau$-scheme \cite{tscheme,Deisz2002,Dee}.

Here we present a scheme to considerably reduce the amount of Matsubara frequencies needed in Eliashberg calculations with Lorentzian-shaped interaction kernel. Our derivations are based on the assumption that the system can be approximated as non-interacting for sufficiently large energies that would correspond to regions of Matsubara space not accessible due to computational limitations. Our method allows for incorporating such infinite tail contributions as corrections to the usually employed hard cut-off scheme without performing any operation in the imaginary time domain. We shall refer to our method as the AT-scheme (analytic tail). In Section \ref{isosec} we introduce the main idea of the AT-scheme using for simplicity the isotropic Eliashberg theory. We briefly discuss possible implicit accuracy improvements of the method on calculating several thermodynamic quantities in the next Section \ref{thermsec}. 
We subsequently proceed with a more complex set of equations in Section \ref{anisosec}, namely the full-bandwidth, multiband and anisotropic Eliashberg equations. For the latter, we choose input parameters that describe superconductivity in the FeSe monolayer on SrTiO$_3$ (STO) substrate, since within this choice the solution to the respective Eliashberg equations is well understood; see Refs.\,\cite{aperis2018,schrodi2018} for details and discussions about various properties of this system in the normal and superconducting state. 
Another reason for choosing FeSe/STO as a case study is that for this system the application of an anisotropic Eliashberg theory that is not restricted to the Fermi level is necessary in order to explain the experimentally observed replica bands \cite{lee2014}. The new method is benchmarked with calculating the spectral function, which is related to Angular Resolved Photoemission Spectroscopy (ARPES) experiments. Details on how to embed our AT-scheme in the self-consistent analytic continuation procedure are provided in  Appendix \ref{appendixac}, and in Appendix \ref{appendixcomp}  we also show the calculated gap on the Fermi surface of the electronic tight-binding model employed for our calculations. In Section \ref{results} we test our proposed method, focusing on the convergence behavior with the number of Matsubara frequencies used.  

For the anisotropic, full-bandwidth and multiband Eliashberg theory we report obtaining well converged results with only one fifth of the number of Matsubara frequencies, compared to the commonly used practice. In the isotropic case, which can easily be generalized to the Fermi surface restricted anisotropic theory \cite{Allen1983}, we spare as many as 9/10 of the frequencies without any decrease in the accuracy of the results. Such a large increase in performance opens up new possibilities for studying low temperature properties within Eliashberg theory and make calculations for very low $T_c$ accessible.

\section{Methodology\label{method}}

\subsection{Isotropic Eliashberg theory\label{isosec}}
As a toy model to introduce our new method we employ the isotropic Eliashberg equations,
\begin{align}
Z_m &= 1 + \frac{\pi T}{\omega_m}\sum_{m^{\prime}} \frac{V_{m-m^{\prime}}\omega_{m^{\prime}}}{\sqrt{\omega_{m^{\prime}}^2+ \Delta^2_{m^{\prime}}}}  , \label{eliashisoz} \\
\Delta_m &=  \frac{\pi T}{Z_m}\sum_{m^{\prime}} \frac{\left[V_{m-m^{\prime}}-\mu^*\right]\Delta_{m^{\prime}}}{\sqrt{\omega_{m^{\prime}}^2+ \Delta^2_{m^{\prime}}}} ~. \label{eliashisod}
\end{align}
Here $Z_m$ is the mass renormalization function and $\Delta_m$ the gap function.
For a system at temperature $T$, $\omega_m=\pi T(2m+1)$ are the fermionic Matsubara frequencies, $m\in\mathbb{Z}$. Throughout this work, sums of the form $\sum_m$ are over an infinite number of Matsubara frequencies, unless noted otherwise. In the simplified case of an Einstein-like phonon mode $\Omega$, we can use a Lorentzian shaped electron-phonon interaction
\begin{align}
V_{m-m^{\prime}} = \frac{\lambda\Omega^2}{\Omega^2 + (\omega_m-\omega_{m^{\prime}})^2}  \label{eliashpot}
\end{align}
to calculate the mass renormalization function $Z_m$ and the gap function $\Delta_m$. In Eq.\,(\ref{eliashpot}), $\lambda$ describes the electron-phonon coupling strength. The renormalized Coulomb interaction enters in Eq.\,(\ref{eliashisod}) via the Anderson-Morel pseudopotential $\mu^*(\omega_c)$, which comes with a cutoff $\omega_c$. In what follows, we assume for simplicity $\mu^*=0$, but we have carefully checked that for $\mu^*\neq0$ the qualitative results do not change. Due to the functional form of the kernel given in Eq.\,(\ref{eliashpot}), the solutions for $Z_m$ and $\Delta_m$ similarly acquire a characteristic Lorentzian shape. In the limit of an infinitely large Matsubara frequency the mass renormalization therefore approaches unity, while the gap function vanishes: $\underset{m\rightarrow\pm\infty}{\lim}Z_m=1$, $\underset{m\rightarrow\pm\infty}{\Delta_m}=0$. Based on this observation we make the assumption
\begin{align}
\exists \mathcal{M}>>1\in\mathbb{N} : Z_{|m|>\mathcal{M}} = 1 ,~ \Delta_{|m|>\mathcal{M}} = 0 \label{assumption1}
\end{align}
for the infinite tails of the Lorentzian functions, which corresponds to the limit of the system being in the non-interacting state outside the interval $[-\mathcal{M},\mathcal{M}]$. As a comparison, the usual practice of introducing the hard cutoff corresponds to the same assumption, but the infinite tails are neglected altogether. 

Assuming Eq.\,(\ref{assumption1}) to hold, we can split the summations in Eqs.\,(\ref{eliashisoz}) and (\ref{eliashisod}) into an interacting and two non-interacting parts. The first non-interacting part is finite for $|m'|\leq\mathcal{M}$ and the second one for $m'\in\mathbb{Z}$. In the following, we label these three terms with $(I)$, $(N_\mathcal{M})$ and $(N)$, respectively. After rewriting the summation over the non-interacting expressions via $\sum_{|m^{\prime}|>\mathcal{M}}=\sum_{m^{\prime}}-\sum_{|m^{\prime}|\leq\mathcal{M}}$ we arrive at an approximated set of Eliashberg equations, labeled by superscript ($\mathcal{A}$):
\begin{align}
Z^{(\mathcal{A})}_m &= 1 + Z^{(I)}_m - Z^{(N_\mathcal{M})}_m + Z^{(N)}_m , \label{isozapprox} \\
\Delta^{(\mathcal{A})}_m &= \Delta^{(I)}_m \label{isodapprox}
\end{align}
Except from the term $Z_m^{(N)}$ in Eq.\,(\ref{isozapprox}), the expressions are easily identified as
\begin{align}
Z^{(I)}_m &=  \frac{\pi T}{\omega_m}\sum_{|m^{\prime}|\leq\mathcal{M}} \frac{V_{m-m^{\prime}}\omega_{m^{\prime}}}{\sqrt{\omega_{m^{\prime}}^2+ \Delta^2_{m^{\prime}}}} , \label{isozs} \\
\Delta^{(I)}_m &=  \frac{\pi T}{Z_m}\sum_{|m^{\prime}|\leq\mathcal{M}} \frac{\left[V_{m-m^{\prime}}-\mu^*\right]\Delta_{m^{\prime}}}{\sqrt{\omega_{m^{\prime}}^2+ \Delta^2_{m^{\prime}}}} , \\
Z^{(N_\mathcal{M})}_m &=  \frac{\pi}{\omega_m}\sum_{|m^{\prime}|\leq\mathcal{M}}V_{m-m^{\prime}} \frac{\omega_{m^{\prime}}}{|\omega_{m^{\prime}}|} ~~. \label{isozn}
\end{align}
The infinite summation occurring in the calculation of $Z_m^{(N)}$ must be treated with special care due to the appearance of a sign function, $\omega_m/ | \omega_m |$. By using the bosonic frequencies $q_m=\omega_m-\pi T=2\pi Tm$ and the complex Digamma function $\psi^{(0)}$ we 
obtain the result as
\begin{align}
&Z^{(N)}_m = \frac{\lambda\Omega}{2\omega_m}\bigg[ \frac{4\pi T\Omega}{q_m^2+\Omega^2} + \pi\coth\left(\frac{\Omega}{2T}\right) ~~~ \\
& ~~~~~~ - i\psi^{(0)}\left(\frac{q_m-i\Omega}{2\pi T}\right) + i\psi^{(0)}\left(\frac{q_m+i\Omega}{2\pi T}\right) \bigg] ~, \nonumber
\end{align}
which together with Eqs.\,(\ref{isozs}-\ref{isozn}) concludes the calculation of $Z_m^{\mathcal{(A)}}$ and $\Delta_m^{\mathcal{(A)}}$. As a consistency check for our derivation we note, that the mass renormalization of Eq.\,(\ref{isozapprox}) in the limit of the non-superconducting state reduces to $Z^{(\mathcal{A})}_m\rightarrow1+\lambda$. The equations presented in this section can straightforwardly be generalized to the anisotropic Eliashberg theory, restricted to momenta around the Fermi level \cite{Allen1983,Choi2002,Margine2013,Aperis2015}.

\subsection{Thermodynamic properties}\label{thermsec}

The gap function $\Delta_m$ and the mass renormalization $Z_m$ as calculated in the previous Section \ref{isosec} can be used to obtain the difference between the normal and superconducting state free energy $F_N-F_S$, which reflects the fact that the superconducting phase is energetically favorable for temperatures $T< T_c$. The expression for the free energy difference $\Delta F$, normalized by the electronic density of states at the Fermi level, reads \cite{carbotte1990},
\begin{align}
&\Delta F(T) = \pi T\sum_m \Bigg[ |\omega_m|\left(Z_m^{(\mathcal{A}_N)}+1\right)   \label{deltaf} \\
&  - 2Z_m^{(\mathcal{A}_S)}\sqrt{\omega_m^2+\Delta_m^2}   + \frac{\omega_m^2\left(Z_m^{(\mathcal{A}_S)}-1\right)+Z_m^{(\mathcal{A}_S)}\Delta_m^2}{\sqrt{\omega_m^2+\Delta_m^2}} \Bigg] ,
\nonumber
\end{align} 
with $Z_m^{(\mathcal{A}_N)}$ and $Z_m^{(\mathcal{A}_S)}$ calculated from Eq.\,(\ref{isozapprox}) for the non-superconducting and superconducting case, respectively. We find the corresponding entropy difference as
\begin{align}
\Delta S = \frac{\partial (\Delta F)}{\partial T} , \label{deltas}
\end{align}
from which we can easily determine the specific heat jump, which is directly observable in experiment and therefore particularly interesting.
\begin{align}
-\frac{\Delta C}{T} = -\frac{1}{T}\frac{\partial (\Delta S)}{\partial T} = -\frac{1}{T}\frac{\partial^2(\Delta F)}{\partial T^2} \label{deltac} .
\end{align}

It is directly evident from Eq.\,(\ref{deltaf}) that our proposed AT-scheme is not applicable for the particular Matsubara sum, since the tails due to the non-superconducting state vanish identically within the approximation made in Eq.\,(\ref{assumption1}). However, due to the fact that within our method we keep the infinite sum over $m'$ in the convolution of Eq.\,(\ref{eliashisoz}), the mass renormalization term entering Eq.\,(\ref{deltaf}) includes such an infinite sum correction. Therefore Eq.\,(\ref{deltaf}) inherits an implicit correction term which can be made explicit if we rewrite Eq.\,(\ref{deltaf}) in a form $\Delta F=\Delta F^{(0)}+\Delta F^{(c)}$, where the superscripts $(0),(c)$ denote the usual $\Delta F$ calculated with a hard cutoff scheme and the correction term generated within our AT-scheme, respectively. The correction term adopts the form
\begin{align}\no
\Delta F^{(c)}(T) = \pi T\sum_m \left(Z^{(N)}_m - Z^{(N_\mathcal{M})}_m\right) |\omega_m|\\\label{deltafcor}
\times\Bigg[ 1 - \sqrt{1+\left(\frac{\Delta_m}{\omega_m}\right)^2}\Bigg] ~.
\end{align}

\subsection{Anisotropic, full-bandwidth and multiband Eliashberg theory}\label{anisosec}

The AT-scheme can be straightforwardly extended to cases where less approximate Eliashberg theory equations need to be solved. As a more complex example we treat the full-bandwidth and anisotropic theory describing the FeSe monolayer on STO substrate. From the detailed analysis in Ref.\,\citealp{aperis2018} we consider the equations
\begin{align}
Z_{{\bf k},m} &= 1 + \frac{T}{\omega_m}\sum_{{\bf k}^{\prime},n}\sum_{m^{\prime}}V_{{\bf q},m-m^{\prime}}\frac{\omega_{m^{\prime}}Z_{{\bf k}^{\prime},m^{\prime}}}{\Theta_{n,{\bf k}^{\prime},m^{\prime}}} , \label{anisotz} \\
\chi_{{\bf k},m} &= -T\sum_{{\bf k}^{\prime},n}\sum_{m^{\prime}}V_{{\bf q},m-m^{\prime}}\frac{\beta_{n,{\bf k}'}+\chi_{{\bf k}^{\prime},m^{\prime}}}{\Theta_{n,{\bf k}^{\prime},m^{\prime}}} , \label{anisotchi} \\
\phi_{{\bf k},m} &= T\sum_{{\bf k}^{\prime},n}\sum_{m^{\prime}}V_{{\bf q},m-m^{\prime}}\frac{\phi_{{\bf k}^{\prime},m^{\prime}}}{\Theta_{n,{\bf k}^{\prime},m^{\prime}}} \label{anisotphi} , \\
\Theta_{n,{\bf k},m} &= \left[\omega_mZ_{{\bf k},m}\right]^2 + \phi^2_{{\bf k},m} + \left[ \beta_{n,{\bf k}}+\chi_{{\bf k},m} \right]^2  ,\label{thetadef}
\end{align}
as a given starting point, where $\mathbf{q}=\mathbf{k}-\mathbf{k}'$. The additional function $\chi_{{\bf k},m}$ represents the chemical potential renormalization and increases the number of coupled equations by one. For brevity we define $\beta_{n,{\bf k}}=\xi_{n,{\bf k}}-\mu$, with $\xi_{n,{\bf k}}$ the momentum dependent bare energy dispersion and $n$ the energy band index.
The global chemical potential, $\mu$, is self-consistently determined  so that the converged solutions to Eq.\,(\ref{anisotz}-\ref{anisotphi}) satisfy the conservation law for the number of electrons by keeping the electron filling
\begin{eqnarray}
n_1 &=& 1 + \frac{2T}{L}\sum_{{\bf k}^{\prime},n}\sum_{m^{\prime}} \frac{\beta_{n,{\bf k}'}+\chi_{{\bf k}^{\prime},m^{\prime}}}{\Theta_{n,{\bf k}^{\prime},m^{\prime}}}
\label{n1}
\end{eqnarray}
constant.

The gap function considered in Section \ref{isosec} is connected to $\phi_{{\bf k},m}$ via $\Delta_{{\bf k},m}=\phi_{{\bf k},m}/Z_{{\bf k},m}$.  For the kernel we use again a Lorentzian shape, compare Eq.\,(\ref{eliashpot}), while now the scattering $\lambda_{{\bf q}}$ is chosen as a momentum dependent small-{\bf q} coupling \cite{Varelogiannis1998,Aperis2011}. Accordingly, our non-interacting state approximation Eq.\,(\ref{assumption1}) is generalized to
\begin{align}
&\exists \mathcal{M}>>1\in\mathbb{N} : Z_{{\bf k},|m|>\mathcal{M}} = 1 ~, \nonumber  \\ 
&~~~~~~\chi_{{\bf k},|m|>\mathcal{M}} = \phi_{{\bf k},|m|>\mathcal{M}} = 0 ~\forall{\bf k} ~~.\label{assumption2}
\end{align}
For keeping the expressions more compact, we denote the bare state limit of $\Theta_{n,{\bf k},m}$ in Eq.\,(\ref{thetadef}) by $\Gamma_{n,{\bf k},m}=\omega_m^2 + \beta_{n,{\bf k}}^2$. The derivation followed in Section \ref{isosec} directly leads to the equations,
\begin{align}
Z^{(I)}_{{\bf k},m} &= \frac{T}{\omega_m}\sum_{{\bf k}^{\prime},n}\sum_{|m^{\prime}|\leq \mathcal{M}}V_{{\bf q},m-m^{\prime}}\frac{\omega_{m^{\prime}}Z_{{\bf k}^{\prime},m^{\prime}}}{\Theta_{n,{\bf k}^{\prime},m^{\prime}}} ,  \\
\chi^{(I)}_{{\bf k},m}   &= T\sum_{{\bf k}^{\prime},n}\sum_{|m^{\prime}|\leq \mathcal{M}}V_{{\bf q},m-m^{\prime}}\frac{\beta_{n,{\bf k}^{\prime}}+\chi_{{\bf k}^{\prime},m^{\prime}}}{\Theta_{n,{\bf k}^{\prime},m^{\prime}}} ,  \\
\phi^{(I)}_{{\bf k},m}   &= T\sum_{{\bf k}^{\prime},n}\sum_{|m^{\prime}|\leq \mathcal{M}}V_{{\bf q},m-m^{\prime}}\frac{\phi_{{\bf k}^{\prime},m^{\prime}}}{\Theta_{n,{\bf k}^{\prime},m^{\prime}}}  , \\
Z^{(N_\mathcal{M})}_{{\bf k},m} &= \frac{T}{\omega_m}\sum_{{\bf k}^{\prime},n}\sum_{|m^{\prime}|\leq \mathcal{M}}V_{{\bf q},m-m^{\prime}}\frac{\omega_{m^{\prime}}}{\Gamma_{n,{\bf k}^{\prime},m^{\prime}}} ,  \\
\chi^{(N_\mathcal{M})}_{{\bf k},m}   &= T\sum_{{\bf k}^{\prime},n}\sum_{|m^{\prime}|\leq \mathcal{M}}V_{{\bf q},m-m^{\prime}}\frac{\beta_{n,{\bf k}^{\prime}}}{\Gamma_{n,{\bf k}^{\prime},m^{\prime}}}  ,
\end{align}
that represent the interacting and non-interacting state contributions inside the numerical boundaries $[-\mathcal{M},\mathcal{M}]$. Therefore the mass renormalization within our method can be found as $Z^{(\mathcal{A})}_{{\bf k},m}=1+Z^{(I)}_{{\bf k},m}-Z^{(N_\mathcal{M})}_{{\bf k},m}+Z^{(N)}_{{\bf k},m}$, and similarly $\chi^{(\mathcal{A})}_{{\bf k},m}=- \chi^{(I)}_{{\bf k},m}+\chi^{(N_\mathcal{M})}_{{\bf k},m} - \chi^{(N)}_{{\bf k},m}$ for the chemical potential renormalization, and $\phi^{(\mathcal{A})}_{{\bf k},m}=\phi^{(I)}_{{\bf k},m}$ for the gap function. The expressions involving infinite summations can be written in a closed form by applying the Residue theorem \cite{saff1993},
\begin{align}
Z^{(N)}_{{\bf k},m} &= \sum_{{\bf k}^{\prime},n}\frac{\lambda_{{\bf q}}\Omega^2}{2}\Bigg(\frac{\frac{1}{\Omega}\coth\left(\frac{\Omega}{2T}\right)\left(\Gamma_{n,{\bf k}^{\prime},m}+\Omega^2\right)}{(\omega_m^2+\Omega^2-\beta_{n,{\bf k}^{\prime}}^2)^2+4\omega_m^2\beta_{n,{\bf k}^{\prime}}^2} \nonumber \\
&  - \frac{ 2\beta_{n,{\bf k}^{\prime}}\tanh\left(\frac{\beta_{n,{\bf k}^{\prime}}}{2T}\right)}{(\omega_m^2+\Omega^2-\beta_{n,{\bf k}^{\prime}}^2)^2+4\omega_m^2\beta_{n,{\bf k}^{\prime}}^2} \Bigg)  , \\
\chi^{(N)}_{{\bf k},m} &= \sum_{{\bf k}^{\prime},n}\frac{\lambda_{{\bf q}}\Omega^2}{2} \Bigg( \frac{\frac{\beta_{n,{\bf k}^{\prime}}}{\Omega}\coth\left(\frac{\Omega}{2T}\right)\left(\Gamma_{n,{\bf k}^{\prime},m}-\Omega^2\right)}{(\omega_m^2+\Omega^2-\beta_{n,{\bf k}^{\prime}}^2)^2+4\omega_m^2\beta_{n,{\bf k}^{\prime}}^2} \nonumber \\
& + \frac{\tanh\left(\frac{\beta_{n,{\bf k}^{\prime}}}{2T}\right)\left(\omega_m^2+\Omega^2-\beta_{n,{\bf k}^{\prime}}^2\right)}{(\omega_m^2+\Omega^2-\beta_{n,{\bf k}^{\prime}}^2)^2+4\omega_m^2\beta_{n,{\bf k}^{\prime}}^2} \Bigg) .
\end{align}
From recent investigations performed on this material it is evident that a reliable analytic continuation from the Matsubara to the real frequency axis is necessary to predict and explain outcomes of spectroscopy experiments \cite{aperis2018,schrodi2018}. Therefore we show in \ref{appendixac} how our method can be applied to a self-consistent continuation procedure for obtaining the real frequency results $Z^{(\mathcal{A})}_{{\bf k},\omega}$, $\chi^{(\mathcal{A})}_{{\bf k},\omega}$ and $\phi^{(\mathcal{A})}_{{\bf k},\omega}$. The derivations do, however, not differ much from what we discuss in this section.

The method introduced here for the isotropic and anisotropic Eliashberg theory has been included in the Uppsala Superconductivity (UppSC) code \cite{Aperis2015,bekaert2018,aperis2018,UppSC}.

\section{Results}\label{results}

\subsection{Isotropic Eliashberg theory}

\begin{figure}[t!]
	%	\vspace*{0.2cm}
	\centering
	\begin{overpic}[width = 1.0\columnwidth, clip, unit=1pt]{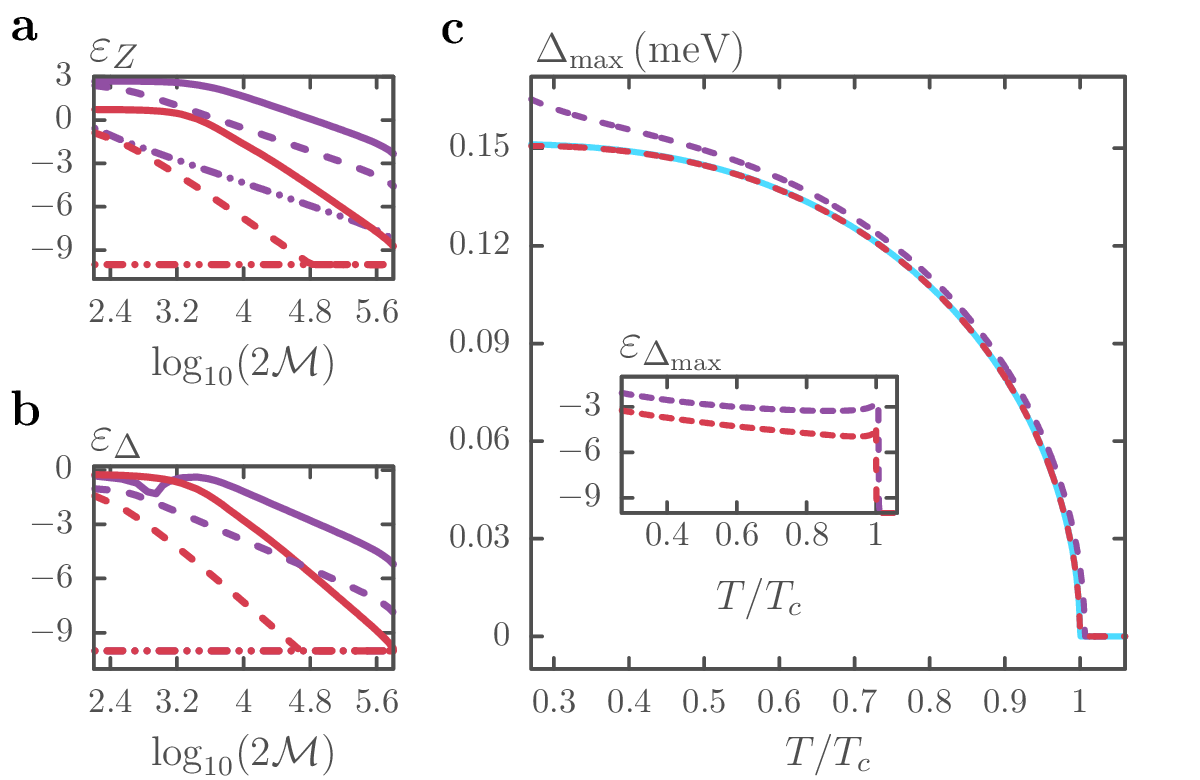}
	\end{overpic}
	%	\vspace*{-0.4cm}
	\caption{Convergence study with the number of Matsubara frequencies used for the self-consistent solution of isotropic Eliashberg equations. Purple colored lines correspond to solutions of the Eqs.\,(\ref{eliashisoz}) and (\ref{eliashisod}), red colored curves to our method for $Z_m^{(\mathcal{A})}$ and $\Delta_m^{(\mathcal{A})}$. Dashed: $\lambda=0.3$, $T=0.1$\,K. Dot-dashed: $\lambda=0.5$, $T=10$\,K. (a), (b) Double logarithmic plot of the average errors $\varepsilon_Z$ and $\varepsilon_{\Delta}$, as defined in the main text, for the mass renormalization and the gap function, respectively. The reference results, calculated for $\mathcal{M}=400\,000$, are considered in an energy window of 200 Matsubara frequencies, centered around the zero. Solid lines correspond to $\lambda=0.5$ and $T=0.01$\,K. (c) Maximum gap as a function of temperature, normalized with respect to $T_c$. The inset shows the logarithmic deviation of $\Delta_{\rm max}$, calculated with $\mathcal{M}=200$, with respect to the cyan solid reference curve obtained for 4\,000 frequencies.}
	\label{isotropic}
\end{figure}

We start by presenting numerical results within the isotropic theory of Section \ref{isosec}. A phonon mode of $\Omega=100$\,K is chosen in all simulations of Eqs.\,(\ref{eliashisoz}) and (\ref{eliashisod}). In Fig.\,\ref{isotropic} the convergence behavior with respect to the cutoff variable $\mathcal{M}$ of the solution to the non-modified equations is compared with the respective solution to the expressions (\ref{isozapprox}) and (\ref{isodapprox}) for $Z_m^{(\mathcal{A})}$ and $\Delta_m^{(\mathcal{A})}$.

In panel Fig.\,\ref{isotropic}(a) we show the logarithmic averaged deviation of the mass renormalization from a reference value $Z_{\text{ref}}$, that is calculated for the very large frequency cutoff $\mathcal{M}=400\,000$. The deviation (i.e., error) is defined as $\varepsilon_Z=\log_{10}\left(\langle|Z_{\text{ref}}-Z_m|\rangle_{|m|\leq100}\right)$ within a window of 200 Matsubara frequencies centered around zero. Figure\,\ref{isotropic}(b) shows the corresponding results for the gap function, where we have defined the error $\varepsilon_{\Delta}$ similarly to $\varepsilon_Z$. The purple colored lines (full, dashed and dashed-dotted) represent the results from Eqs.\,(\ref{eliashisoz}) and (\ref{eliashisod}) where a hard cutoff has been applied, while with red lines are shown the results after applying our AT-scheme, Eqs.\,(\ref{isozapprox}) and (\ref{isodapprox}). The dot-dashed lines are found by using the coupling $\lambda=0.5$ at a temperature $T=10$\,K $>T_c$. 
Since for these parameters of our toy-model, the system is not superconducting, $\Delta_m=\Delta_m^{(\mathcal{A})}=0$ and $\varepsilon_\Delta$ is infinitesimal. In Fig.\,\ref{isotropic} the latter value coincides with our chosen numerical resolution of $10^{-10}$. Note, that the error for the mass renormalization remains finite in the case of the hard cutoff scheme while it vanishes for our AT-scheme.
The solid lines in these panels are calculated for $\lambda=0.5$ and $T=0.01$\,K$<T_c$. In such a low-$T$ simulation it is much more difficult to achieve convergence. It is clearly revealed, that our method reaches the same accuracy with about one order of magnitude less Matsubara frequencies. Equivalently, for a fixed value of the cutoff $\mathcal{M}$ we report an error about three orders of magnitude smaller.

The dashed results in each panel of Fig.\,\ref{isotropic} were obtained with a coupling of $\lambda=0.5$; for panels (a) and (b) we used $T=0.1$\,K and again find a remarkable increase in performance. To estimate this behavior for an observable quantity, we calculated the maximum gap value as a function of temperature. 
Panel (c) shows results for the same coupling strength, $\lambda=0.5$, and within a chosen cutoff $\mathcal{M}=100$ for which our method yields an error of $\varepsilon_\Delta=10^{-6}$.
The cyan reference curve $\Delta_{\text{max}}^{\text{ref}}$ is obtained from Eqs.\,(\ref{eliashisoz}) and (\ref{eliashisod}) for 4000 Matsubara frequencies. As revealed by the logarithmic error $\varepsilon_{\Delta_{\text{max}}}=\log_{10}(|\Delta_{\text{max}}^{\text{ref}}-\Delta_{\text{max}}|)$ in the inset, we can reliably reproduce the reference curve at least up to $\mu$eV precision, while the non-corrected calculations are more than one order of magnitude less accurate. The low temperature results we obtain are significantly better than the non-corrected calculations, which even become unphysical at small $T$, since a BCS behavior is to be expected for such a moderate interaction strength.

\subsection{Thermodynamic properties}
\begin{figure}[ht!]
	%	\vspace*{0.8cm}
	\centering
	\begin{overpic}[width = 1.0\columnwidth, clip, unit=1pt]{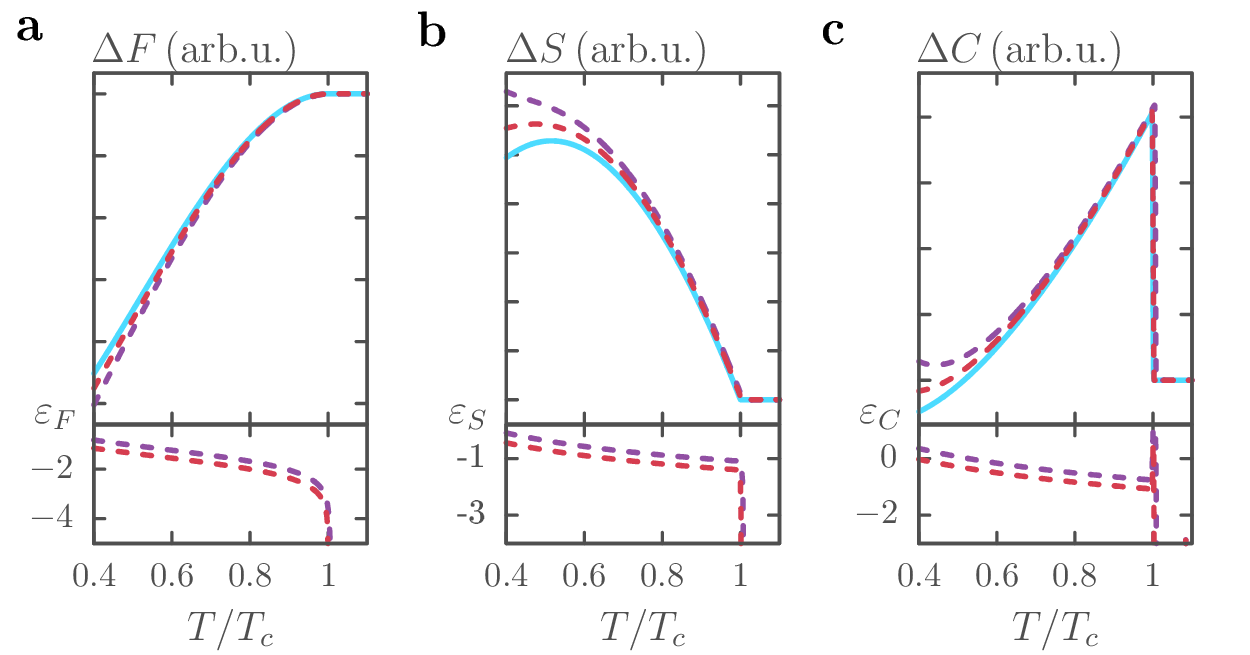}
	\end{overpic}
	%	\vspace*{-0.4cm}
	\caption{Thermodynamic properties obtained from the solution of the isotropic Eliashberg equations for a coupling strength $\lambda=0.3$. Beneath each panel we show the logarithmic error with respect to the reference curve. The color code and the number of Matsubara frequencies is chosen precisely as in Fig.\,\ref{isotropic}(c). (a) Free energy difference Eq.\,(\ref{deltaf}) between normal and superconducting state. (b) Difference in entropy, Eq.\,(\ref{deltas}). (c) Specific heat difference [Eq.\,(\ref{deltac})] that shows the characteristic jump at $T_c$.}
	\label{thermo}
\end{figure}

As described in Section \ref{thermsec} we can use the results for the mass renormalization and the gap function to calculate the temperature dependent free energy of the system, making also other thermodynamic quantities accessible; the calculated free energy, entropy and specific heat differences are shown in Fig.\,\ref{thermo}.
Using the same coupling $\lambda=0.3$ and equivalent number of Matsubara frequencies as in Fig.\,\ref{isotropic}, the panels (a), (b) and (c) correspond to the solution of Eqs.\,(\ref{deltaf}), (\ref{deltas}) and (\ref{deltac}), respectively. In addition we show for each panel the logarithmic error $\varepsilon_F=\log_{10}\left(|\Delta F_{\text{ref}}-\Delta F|\right)$ as a function of temperature, with similar definitions for the entropy and the specific heat jump. The solid reference curves (cyan color) are found by solving Eqs.\,(\ref{eliashisoz}) and (\ref{eliashisod}) for ten times more frequencies than for the dashed results.

A close inspection of Fig.\,\ref{thermo} reveals a smaller gain due to our proposed method, comparing with the solution of the Eliashberg equations themselves. As discussed earlier, this ineffectiveness can be explained by the way the free energy difference is calculated.
In Eq.\,(\ref{deltaf}) we need to evaluate a sum over Matsubara frequencies for which the infinite tails cannot be taken into account analytically. Another source of numerical inaccuracies is the precision up to which $Z_m$ and $\Delta_m$ are converged. As shown previously (see, e.g.\ Fig.\,\ref{isotropic}), our AT-scheme is superior to the calculation, in which the tails are neglected. Any accuracy improvements thus propagate into the solution of the thermodynamic properties and approximately double the precision of the latter calculations. This mechanism is particularly effective in the low temperature regime. 

\begin{figure}[t!]
	%	\vspace*{0.2cm}
	\centering
	\begin{overpic}[width = 1.0\columnwidth, clip, unit=1pt]{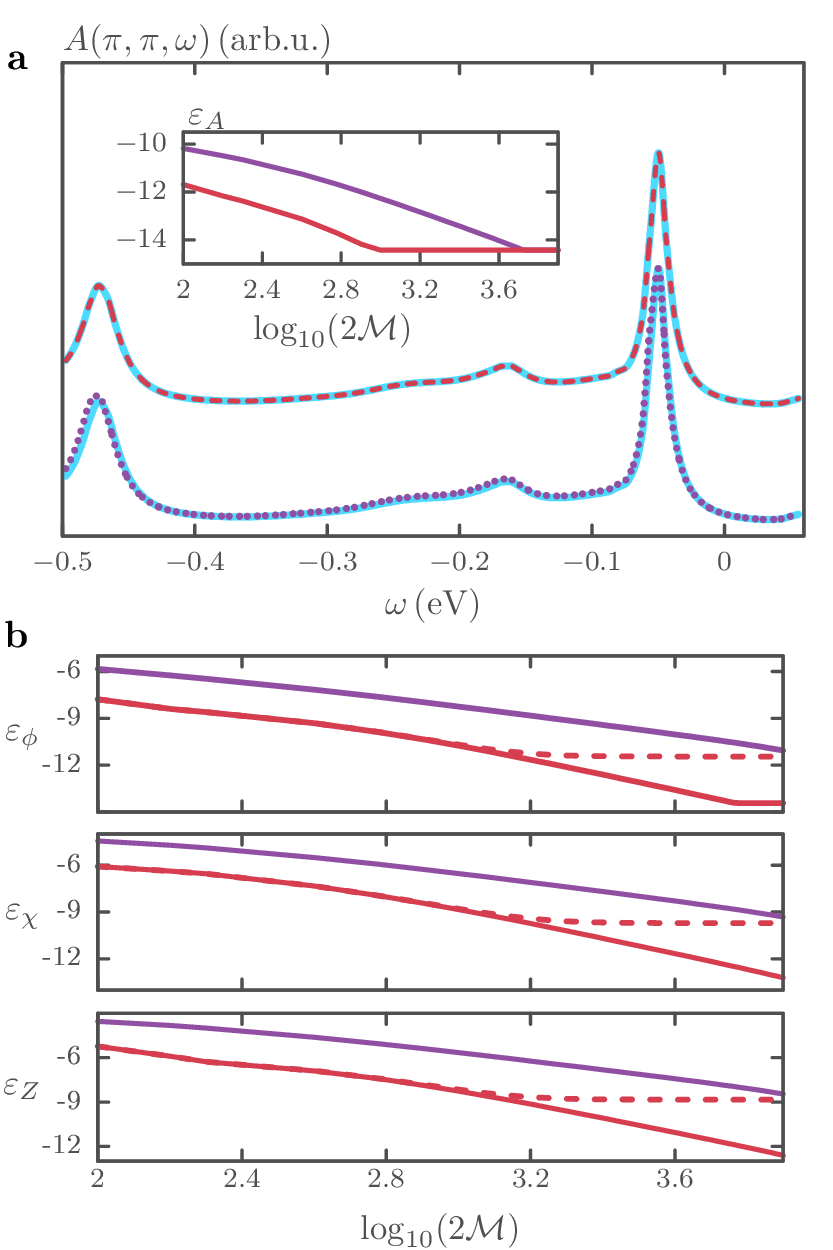}
	\end{overpic}
	%	\vspace*{-0.35cm}
	\caption{Convergence study of the full-bandwidth, multiband and anisotropic Eliashberg equations for the FeSe/STO system at $T=10$\,K. (a) 
		Computed ARPES spectrum at the $M$-point of the first Brillouin zone; the solid and cyan curves correspond to the analytically continued results for $Z_{{\bf k},m}$, $\phi_{{\bf k},m}$ and $\chi_{{\bf k},m}$ for 12\,000 Matsubara frequencies (both lines are similar, but shifted relative to each other for visibility purpose). Purple and dotted (red and dashed) curves: Results for 100 Matsubara frequencies taking (not taking) into account the infinite tail correction. The inset shows the corresponding logarithmic error with respect to the reference line. (b) Logarithmic average error of the real frequency dependent functions as defined in the main text, using the same color code. Dashed: Analytically continued solutions for $Z^{(\mathcal{A})}_{{\bf k},m}$, $\chi^{(\mathcal{A})}_{{\bf k},m}$ and $\phi^{(\mathcal{A})}_{{\bf k},m}$ compared to the reference solutions obtained from Eqs.\,(\ref{anisotz}-\ref{anisotphi}).}
	\vspace*{-0.45cm}
	\label{arpes}
\end{figure}

\subsection{Anisotropic, full-bandwidth and multiband Eliashberg theory}

Let us now turn to a more complex application, namely the full-bandwidth, multiband and anisotropic Eliashberg theory for the FeSe monolayer on STO. For the physical understanding and the derivation of the equations used we refer to previous work on this material\cite{aperis2018,schrodi2018}. Here we put the focus on the convergence behavior with respect to the number of frequencies used in both, the corrected and non-corrected set of Eliashberg equations. Considering the system at temperature $T=10$\,K we start from the very low number of 100 Matsubara frequencies, and subsequently increase the cutoff $\mathcal{M}$, up till 12\,000 frequencies. As a comparison, it has been shown that a choice $\mathcal{M}=1\,500$ is already sufficient to reliably reproduce experimental results for this specific system \cite{aperis2018,schrodi2018}. The functions corresponding to the largest number of Matsubara frequencies again serve as reference values for calculating the errors and are labeled $Z_{\text{ref}}$, $\chi_{\text{ref}}$ and $\phi_{\text{ref}}$ (similarly $Z^{(\mathcal{A})}_{\text{ref}}$, $\chi^{(\mathcal{A})}_{\text{ref}}$ and $\phi^{(\mathcal{A})}_{\text{ref}}$). 

In Fig.\,\ref{arpes}(a) we show the result for the spectral function at the high-symmetry $M$-point of the reduced Brillouin zone, which reveals well documented and material-characteristic features, that are experimentally accessible via ARPES \cite{zhang2017,lee2014,huang2017}.
The two cyan solid reference curves are identical, though shifted relative to each other, and are calculated by self-consistent analytic continuation of Eqs.\,(\ref{anisotz}-\ref{anisotphi}) (see Appendices \ref{appendixac} and \ref{appendixcomp}). The purple dotted lines correspond to neglecting the infinite Matsubara tails, while the red dashed curves represent our proposed method; both are obtained for 100 frequencies. Due to smearing effects and the strict convergence criteria we impose in the self-consistency loop of the analytic continuation procedure both results are remarkably accurate. We note, however, that a visible difference with respect to the reference lines is only present for the non-corrected way of solving the equations. For a more quantitative error measure we show in the inset the logarithmic deviation integrated over frequency, $\varepsilon_A=\log_{10}\left(\int|A-A_{\text{ref}}|\text{d}\omega\right)$ as a function of the frequency cutoff $\mathcal{M}$ using the same color code. The reference value is denoted by  $A_{\text{ref}}$ and similarly for $A^{(\mathcal{A})}$. We report a gain of at least a factor five for the number of Matsubara frequencies needed to reach a given precision. Correspondingly, for a given cutoff $\mathcal{M}$ our method produces an error two orders of magnitude smaller. Figure\,\ref{arpes}(b) shows the convergence study with respect to the analytically continued $Z_{{\bf k},\omega}$, $\chi_{{\bf k},\omega}$ and $\phi_{{\bf k},\omega}$, as defined in Appendix \ref{appendixac}. By using the same cutoff for the reference solutions $Z_{\text{ref}}$, $\chi_{\text{ref}}$ and $\phi_{\text{ref}}$ as for the spectral function, we define the errors as $\varepsilon_Z=\log_{10}\left(\langle |Z_{\text{ref}}-Z_{{\bf k},\omega}|\rangle_{{\bf k},\omega}\right)$, $\varepsilon_{\chi}=\log_{10}\left(\langle |\chi_{\text{ref}}-\chi_{{\bf k},\omega}|\rangle_{{\bf k},\omega}\right)$ and $\varepsilon_{\phi}=\log_{10}\left(\langle |\phi_{\text{ref}}-\phi_{{\bf k},\omega}|\rangle_{{\bf k},\omega}\right)$, and similar for our corrected results $(\mathcal{A})$. For these quantities we observe, that the savings in performance are even larger, and up to a factor of eight in the number of frequencies. For completeness we added a comparison of our method with the reference values for the non-corrected procedure, shown by the dashed red curves, from which we can draw the same conclusions as already mentioned above.

\section{Concluding remarks}

In this work we have proposed a method (the AT-scheme) for taking into account analytically the infinite tails of Matsubara frequency dependent functions when performing Matsubara frequency sums in self-consistent Eliashberg calculations with Lorentzian shaped kernel. The application of our AT-scheme was demonstrated to lead to a considerable increase in computational performance. Benchmarking our method within the isotropic Eliashberg theory reveals that the convergence with respect to the number of frequencies is faster and smoother compared to using a hard cutoff scheme. For the isotropic case we spare 90\% and for the anisotropic equations 80\% in the number of Matsubara frequencies needed to reach convergence with high numerical accuracy. The calculation of thermodynamic quantities involves a second infinite summation, that cannot directly be treated within the formalism we introduce. This leads to a smaller gain in numerical savings, though we still outperform the direct implementation by a factor of two. We note that the increase in efficiency due to our here-presented corrections does not change with system characteristics, like the phonon frequency or coupling strength.

The $\tau$-scheme provides a very efficient way of calculating Matsubara frequency sums for arbitrary-shaped interaction kernels (as recently demonstrated in Ref.\ \onlinecite{Dee}). 
However, the application of this approach has thus far been limited to calculations that involve off-diagonal self-energies (e.g. superconducting) in the linearized regime. Whether the $\tau$-scheme can be implemented below $T=T_c$ is an open problem. On the other hand, our proposed AT-scheme is applicable regardless the temperature and it, in fact, may as well become
a necessity for $T\rightarrow 0$, since then the required number of Matsubara frequencies is very large. Perhaps the trade-off for the otherwise wide applicability of our AT-scheme is the fact that it requires interactions with Lorentzian-shaped kernels which is, nevertheless, a quite common physical situation. Whether this assumption can be relaxed to arbitrary-shaped kernels, e.g.\ by means of suitable interpolation/extrapolation procedures, is an interesting question for future investigation.

Our results are representative for coupled sets of equations within the Migdal-Eliashberg formalism, regardless of the degree to which these are simplified. Other physical situations, e.g. including an external magnetic field \cite{Aperis2015}, can easily be considered, since the structure of the describing equations does not change. Further we stress, that our formalism works for even- and odd-frequency pairing channels, and that there is no restriction on the superconductivity mediating mechanism. The procedure we discuss here has numerous applications in explaining experimentally observed phenomena in superconductors at low temperature. The recent discoveries of transition temperatures as low as $T_c=1.5$\,K in bilayer graphene \cite{cao2018}, $T_c=1.8$\,K in the chalcogenide Nb$_2$PdS$_5$ \cite{park2018} or $T_c=0.38$\,K in PdTe$_2$ \cite{das2018} are only a few examples for the wide applicability of our method.

\section*{Acknowledgements}
\noindent
This work has been supported by the Swedish Research Council (VR), the R{\"o}ntgen-{\AA}ngstr{\"o}m Cluster, and the Swedish National Infrastructure for Computing (SNIC).

\appendix

\section{Analytic continuation} \label{appendixac}

For obtaining real frequency dependent results from the functions found in Section \ref{anisosec} we follow a self-consistent algorithm, that is formally exact \cite{marsiglio1988}. Since we want to focus on the implications of assumption (\ref{assumption2}) on the corresponding equations, we refer to Refs.\,\cite{aperis2018,schrodi2018} for further details about the implementation and derivation. We start by writing the equations given in Ref.\,\cite{aperis2018} in a form suitable for our purpose:
\begin{align}
&Z_{{\bf k},\omega} = Z^{(L)}_{{\bf k},\omega} + \frac{1}{2\omega} \int_{-\infty}^{\infty}\text{d}z^{\prime} \sum_{{\bf k}^{\prime},n}\tilde{\alpha}F({\bf k},{\bf k}^{\prime},z^{\prime}) \label{analyticfullz} \\
& \cdot \left[\tanh\left(\frac{\omega-z^{\prime}}{2T}\right)+\coth\left(\frac{z^{\prime}}{2T}\right)\right] \frac{Z_{{\bf k}^{\prime},\omega-z^{\prime}}(\omega-z^{\prime})}{\Theta_{n,{\bf k}^{\prime},\omega-z^{\prime}}}, \nonumber \\
&\chi_{{\bf k},\omega} = \chi^{(L)}_{{\bf k},\omega} - \frac{1}{2} \int_{-\infty}^{\infty}\text{d}z^{\prime} \sum_{{\bf k}^{\prime},n}\tilde{\alpha}F({\bf k}, {\bf k}^{\prime},z^{\prime})  \label{analyticfullc} \\
& \cdot\left[\tanh\left(\frac{\omega-z^{\prime}}{2T}\right)+\coth\left(\frac{z^{\prime}}{2T}\right)\right] \frac{\beta_{n,{\bf k}^{\prime}} + \chi_{{\bf k}^{\prime},\omega-z^{\prime}}}{\Theta_{n,{\bf k}^{\prime},\omega-z^{\prime}}} , \nonumber \\
&\phi_{{\bf k},\omega} = \phi^{(L)}_{{\bf k},\omega} + \frac{1}{2} \int_{-\infty}^{\infty}\text{d}z^{\prime} \sum_{{\bf k}^{\prime},n}\tilde{\alpha}F({\bf k}, {\bf k}^{\prime},z^{\prime}) \label{analyticfullp} \\
& \cdot \left[\tanh\left(\frac{\omega-z^{\prime}}{2T}\right)+\coth\left(\frac{z^{\prime}}{2T}\right)\right] \frac{\phi_{{\bf k}^{\prime},\omega-z^{\prime}}} {\Theta_{n,{\bf k}^{\prime},\omega-z^{\prime}}} . \nonumber 
\end{align}
Given in this way, only the first term on each right hand side of Eqs.\,(\ref{analyticfullz}-\ref{analyticfullp}) depends on the results in Matsubara space, therefore we focus on these expressions. By using the Lorentzian-shaped electron-phonon interaction $V_{{\bf q}}(\omega-\omega_{m^{\prime}})$ as introduced in the main text, we identify
\begin{align}
Z^{(L)}_{{\bf k},\omega} &= 1 + \frac{T}{\omega} \sum_{{\bf k}^{\prime},n}\sum_{m^{\prime}} \frac{V_{{\bf q}}(\omega-\omega_{m^{\prime}})Z_{{\bf k}^{\prime},m^{\prime}}i\omega_{m^{\prime}}}{\Theta_{n,{\bf k}^{\prime},m^{\prime}}}  , \\ 
\chi^{(L)}_{{\bf k},\omega} &= - T \sum_{{\bf k}^{\prime},n}\sum_{m^{\prime}} \frac{V_{{\bf q}}(\omega-\omega_{m^{\prime}})(\beta_{n,{\bf k}^{\prime}} + \chi_{{\bf k}^{\prime},m^{\prime}})}{\Theta_{n,{\bf k}^{\prime},m^{\prime}}} ,  \\
\phi^{(L)}_{{\bf k},\omega} &= T \sum_{{\bf k}^{\prime},n}\sum_{m^{\prime}} \frac{V_{{\bf q}}(\omega-\omega_{m^{\prime}})\phi_{{\bf k}^{\prime},m^{\prime}}}{\Theta_{n,{\bf k}^{\prime},m^{\prime}}} \label{analyticMat} .
\end{align}
After employing assumption\,(\ref{assumption2}) and rewriting the summations appropriately we find a set of modified equations 
\begin{align}
Z^{(L,\mathcal{A})}_{{\bf k},\omega} &= 1 + Z^{(L,I)}_{{\bf k},\omega} - Z^{(L,N_\mathcal{M})}_{{\bf k},\omega} + Z^{(L,N)}_{{\bf k},\omega} , \label{zl}  \\
\chi^{(L,\mathcal{A})}_{{\bf k},\omega} &= -\chi^{(L,I)}_{{\bf k},\omega} + \chi^{(L,N_\mathcal{M})}_{{\bf k},\omega} - \chi^{(L,N)}_{{\bf k},\omega} ,\label{chl} \\
\phi^{(L,\mathcal{A})}_{{\bf k},\omega} &= \phi^{(L,I)}_{{\bf k},\omega} ,
\end{align}
the constituents of which are similarly defined as in Section \ref{anisosec}. The interacting and non-interacting state terms confined to the Matsubara frequency interval $[-\mathcal{M},\mathcal{M}]$ are easily found as
\begin{align}
Z^{(L,I)}_{{\bf k},\omega} &= \frac{T}{\omega} \sum_{{\bf k}^{\prime},n}\sum_{|m^{\prime}|\leq \mathcal{M}} \frac{V_{{\bf q}}(\omega-\omega_{m^{\prime}})Z_{{\bf k}^{\prime},m^{\prime}}i\omega_{m^{\prime}}}{\Theta_{n,{\bf k}^{\prime},m^{\prime}}} \\
\chi^{(L,I)}_{{\bf k},\omega} &= T \sum_{{\bf k}^{\prime},n}\sum_{|m^{\prime}|\leq \mathcal{M}} \!\!\! \frac{V_{{\bf q}}(\omega-\omega_{m^{\prime}})(\beta_{n,{\bf k}^{\prime}} + \chi_{{\bf k}^{\prime},m^{\prime}})}{\Theta_{n,{\bf k}^{\prime},m^{\prime}}} ~~~~~~ \\
\phi^{(L,I)}_{{\bf k},\omega} &= T \sum_{{\bf k}^{\prime},n}\sum_{|m^{\prime}|\leq \mathcal{M}} \frac{V_{{\bf q}}(\omega-\omega_{m^{\prime}})\phi_{{\bf k}^{\prime},m^{\prime}}}{\Theta_{n,{\bf k}^{\prime},m^{\prime}}} \\
Z^{(L,N_\mathcal{M})}_{{\bf k},\omega} &= \frac{T}{\omega} \sum_{{\bf k}^{\prime},n}\sum_{|m^{\prime}|\leq \mathcal{M}} \frac{V_{{\bf q}}(\omega-\omega_{m^{\prime}})i\omega_{m^{\prime}}}{\Gamma_{n,{\bf k}^{\prime},m^{\prime}}} \\
\chi^{(L,N_\mathcal{M})}_{{\bf k},\omega} &= T \sum_{{\bf k}^{\prime},n}\sum_{|m^{\prime}|\leq \mathcal{M}} \frac{V_{{\bf q}}(\omega-\omega_{m^{\prime}})\beta_{n,{\bf k}^{\prime}} }{\Gamma_{n,{\bf k}^{\prime},m^{\prime}}} ~~,
\end{align}
while the expressions involving infinite summations require more effort to calculate. It is easy to check that the poles of the summands are never given by an integer number, which makes the Residue theorem for infinite summations again applicable \cite{saff1993}. For sake of brevity we define $\omega_+=\omega+\Omega$ and $\omega_-=\omega-\Omega$, which leads to the terms
\begin{align}
&Z_{{\bf k},\omega}^{(L,N)} = \sum_{{\bf k}^{\prime},n}\frac{\lambda_{{\bf q}}\Omega^2}{4\omega}\left[ \frac{\tanh\left(\frac{\omega_-}{2T}\right)}{\omega_-+\beta_{n,{\bf k}^{\prime}}}\frac{\omega_-/\Omega}{\omega_--\beta_{n,{\bf k}^{\prime}}}   \right. \nonumber \\
& ~~~~~~ +  \frac{\tanh\left(\frac{\beta_{n{\bf k}^{\prime}}}{2T}\right)}{\omega_-+\beta_{n,{\bf k}^{\prime}}}\frac{1}{\omega_++\beta_{n,{\bf k}^{\prime}}}  - \frac{\tanh\left(\frac{\beta_{n,{\bf k}^{\prime}}}{2T}\right)}{\omega_--\beta_{n,{\bf k}^{\prime}}} \nonumber \\
& \left. ~~~~~~ \times\frac{1}{\omega_+-\beta_{n,{\bf k}^{\prime}}} - \frac{\tanh\left(\frac{\omega_+}{2T}\right)}{\omega_++\beta_{n,{\bf k}^{\prime}}}\frac{\omega_+/\Omega}{\omega_+-\beta_{n,{\bf k}^{\prime}}}    \right] ,  \\
&\chi_{{\bf k},\omega}^{(L,N)} = \sum_{{\bf k}^{\prime},n}\frac{\lambda_{{\bf q}}\Omega^2}{4}\left[  \frac{\tanh\left(\frac{\omega_-}{2T}\right)}{\omega_-+\beta_{n,{\bf k}^{\prime}}}\frac{\beta_{n,{\bf k}^{\prime}}/\Omega}{\omega_--\beta_{n,{\bf k}^{\prime}}}  \right. \nonumber  \\
& ~~~~~~  - \frac{\tanh\left(\frac{\beta_{n,{\bf k}^{\prime}}}{2T}\right)}{\omega_++\beta_{n,{\bf k}^{\prime}}}\frac{1}{\omega_-+\beta_{n,{\bf k}^{\prime}}} - \frac{\tanh\left(\frac{\beta_{n,{\bf k}^{\prime}}}{2T}\right)}{\omega_+-\beta_{n,{\bf k}^{\prime}}}  \nonumber \\
& \left. ~~~~~~ \times \frac{1}{\omega_--\beta_{n,{\bf k}^{\prime}}} -   \frac{\tanh\left(\frac{\omega_+}{2T}\right)}{\omega_++\beta_{n,{\bf k}^{\prime}}}\frac{\beta_{n,{\bf k}^{\prime}}/\Omega}{\omega_+-\beta_{n,{\bf k}^{\prime}}}    \right]  , 
\end{align}	
that conclude the calculation of Eqs.\,(\ref{zl}) and (\ref{chl}), and therefore determine the results found by the self-consistent Eqs.\,(\ref{analyticfullz}-\ref{analyticfullp}).

\section{Computational Details} \label{appendixcomp}

The isotropic Eliashberg Eqs.\,(\ref{eliashisoz}) and (\ref{eliashisod}) are solved iteratively with a convergence criterion of a maximal error $10^{-10}$ between two successive iterations. The highest number of iterations is chosen as 15\,000. Within our numerical algorithm, all convolutions are performed via the Fast Fourier Transform technique. The free energy difference is obtained by the Matsubara frequency dependent solutions for $Z_m$ and $\Delta_m$ using Eq.\,(\ref{deltaf}). From this we calculate Eq.\,(\ref{deltas}) and (\ref{deltac}), the results of which are given in Fig.\,\ref{thermo}, without numerical smoothing or fitting.

\begin{figure}[t!]
	%	\vspace*{0.2cm}
	\centering
	\begin{overpic}[width = 0.75\columnwidth, clip, unit=1pt]{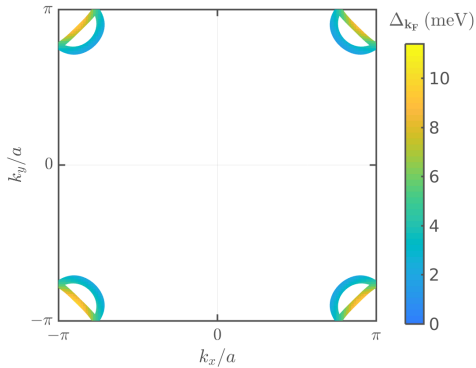}
	\end{overpic}
	%	\vspace*{-0.35cm}
	\caption{Self-consistently obtained superconducting gap of monolayer FeSe/STO projected on the electronic Fermi surface pockets. The gap is determined at $T=10\,\mathrm{K}$ and using a coupling as described in the main text. The size of the band and momentum dependent  gap is depicted by the color code.}
	\vspace*{-0.45cm}
	\label{fs}
\end{figure}

For the anisotropic case discussed in Fig.\,\ref{arpes} we used a $128\times128-{\bf k}$ grid and a total of 5\,000 real frequencies. As explained in more detail in Refs.\,\cite{aperis2018,schrodi2018} we employ a bare ten-band energy dispersion originally obtained from Density Functional Theory calculations. For the specific choice of the constants appearing within this description (global chemical potential, averaged phonon frequency, etc.) we used parameter ranges, that have been proven to give results well in agreement with experiment \cite{aperis2018}. As example of the computed superconducting quantities we show in Fig.\,\ref{fs} the zero-frequency superconducting gap as obtained from our self-consistent Eliashberg calculation, projected on the Fermi surface of the tight-binding model. Our results for the size of the gap compare well to the experimentally measured gap size  \cite{zhang2017}. For both, the Matsubara space calculation and the analytic continuation, we used a minimal error of $10^{-10}$ for convergence and a maximal number of 10\,000 iterations.

\end{document}